\documentclass{emulateapj-rtx4}
\usepackage{apjfonts}

\def\gax{\mathrel{\raise.3ex\hbox{$>$}\mkern-14mu\lower0.6ex\hbox{$\sim$}}}
\def\lax{\mathrel{\raise.3ex\hbox{$<$}\mkern-14mu\lower0.6ex\hbox{$\sim$}}}
\def\gtorder{\mathrel{\raise.3ex\hbox{$>$}\mkern-14mu
             \lower0.6ex\hbox{$\sim$}}}
\def\ltorder{\mathrel{\raise.3ex\hbox{$<$}\mkern-14mu
             \lower0.6ex\hbox{$\sim$}}}

\begin{document}

\title{Dust to Dust: 3 years in the Evolution of the Unusual SN~2008S}

\author{D.~M. Szczygie{\l}$^{1}$, J.~L. Prieto$^{2,7}$,
  C.~S. Kochanek$^{1,4}$, K.~Z. Stanek$^{1,4}$,
  T.~A. Thompson$^{1,4}$, J.~F. Beacom$^{1,3,4}$,
  \newline P.~M. Garnavich$^{5}$, C.~E. Woodward$^{6}$}
 
\altaffiltext{1}{Department of Astronomy, The Ohio State University, 140 W. 18th Ave., Columbus OH 43210}
\altaffiltext{2}{Department of Astrophysical Sciences, Princeton University, Peyton Hall, Princeton, NJ 08544}
\altaffiltext{3}{Department of Physics, The Ohio State University, 191 W. Woodruff Ave., Columbus OH 43210}
\altaffiltext{4}{Center for Cosmology and AstroParticle Physics, The Ohio State University, 191 W. Woodruff Ave., Columbus OH 43210}
\altaffiltext{5}{University of Notre Dame, 225 Nieuwland Science Hall, Notre Dame, IN 46556}
\altaffiltext{6}{ Minnesota Institute for Astrophysics, School of Physics and Astronomy, 116 Church Street, S. E., University of Minnesota, Minneapolis, MN 55455}
\altaffiltext{7}{Hubble and Carnegie-Princeton Fellow}

\begin{abstract}

\noindent We obtained late-time optical and near-IR imaging of
SN~2008S with the Large Binocular Telescope (LBT), near-IR data with
the Hubble Space Telescope (HST), and mid-IR data with the Spitzer
Space Telescope (SST). We find that (1) it is again invisible at optical
($UBVR$) wavelengths to magnitude limits of approximately $25$~mag,
(2) while detected in the near-IR ($H$) at approximately $24.8$~mag,
it is fading rapidly, and (3) it is still brighter than the progenitor
at $3.6$ and $4.5\mu$m in the mid-IR with a slow, steady decline.  The IR
detections in December 2010 are consistent with dust emission at
a blackbody temperature of $T \simeq 640$~K and a total luminosity of
$L \simeq 200000$~$L_\odot$, much higher than the $L\simeq 40000 L_\odot$ 
luminosity of the obscured progenitor star. The local environment
also shows no evidence for massive ($M \gtorder 10 M_\odot$) stars in
the vicinity of the transient, consistent with the progenitor being
a massive AGB star.  
\end{abstract}


\section{Introduction}
\label{sec:introduction}

SN~2008S is one of the most mysterious optical transients created by a
massive star in the last decade. It was discovered in February 2008 by
\cite{Arbour2008} in the prolific supernova factory NGC~6946. It was
initially classified as a likely ``supernova impostor" due to its
faint absolute peak magnitude ($M_V \sim -13$~mag) and optical spectra
dominated by narrow Balmer, Ca~II triplet, and [Ca~II] lines in emission
(\citealt{Stanishev2008}; \citealt{Steele2008}). NGC~6946 had been
observed by the Large Binocular Telescope (LBT) the previous year, and
the key piece of evidence from these observations was that there was
no optical progenitor \citep{Prieto2008}, which was surprising since
the ``supernova impostors'' are believed to be eruptions from very
massive ($>20$-$30\,M_\odot$), evolved stars (e.g.,
\citealt{Smith2010} and references therein) that should have been
easily visible in the LBT observations.

\begin{figure*}
\centerline{\includegraphics[width=18cm]{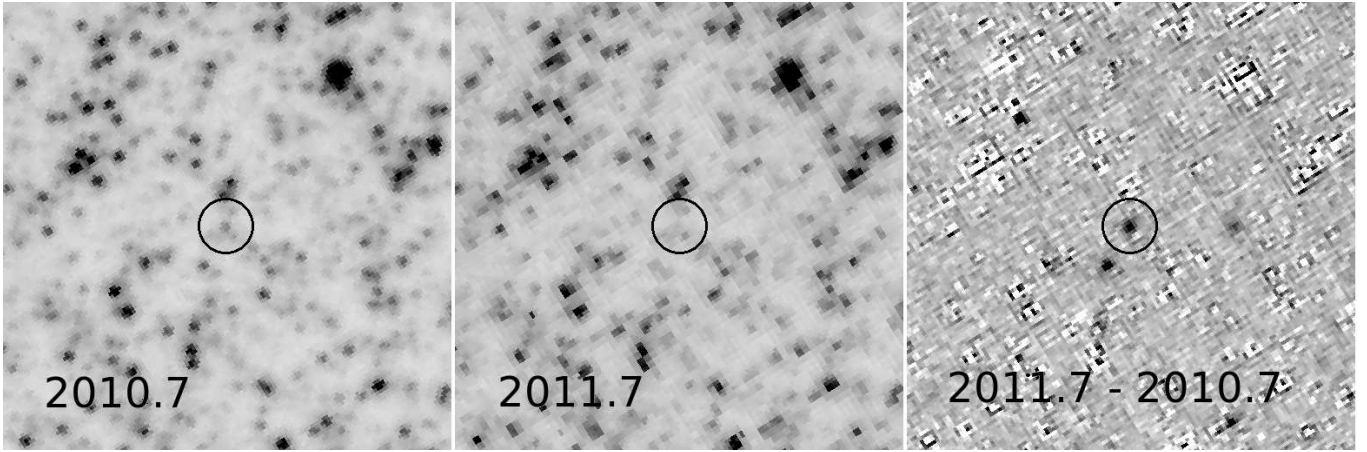}}
\hspace{0.15cm}
\caption{The HST F160W ($H$) band observations of SN~2008S from August
2010 (left) and 2011 (middle). The right panel shows the difference
between the 2011 and 2010 epochs, where black means that the source
has become fainter. 
There are other variables in the field of view, but with only two
epochs we cannot classify them.  There are no ambiguities in the
astrometry on the scale of the distance to the next nearest variable
source.  The panels are 8\farcs2 $\times$ 8\farcs2
($200 \times 200$ pc) and the radius of the circle is 0\farcs5.}
\label{fig:hst}
\end{figure*}

\begin{table*}
\tabletypesize{\scriptsize}
\begin{center}
\caption{Photometry of SN~2008S \label{tab:magnitudes}}
\begin{tabular}{ccccccccccc}
\tableline
Date & MJD & $Us$ & $B$ & $V$ & $R$ & $J$ & $H$ & $K$ & $[3.6]$ & $[4.5]$ \\
(UT) &  & (mag) & (mag) & (mag) & (mag) & (mag) & (mag) & (mag) & (mag) & (mag) \\
\tableline
2008-02-06 & 54503.3  &$\cdots$       &$\cdots$        &$\cdots$       &$\cdots$       &$\cdots$       &$\cdots$       &$\cdots$       &$13.11\pm0.01$ &$12.40\pm0.01$\\
2008-05-03 & 54589.4  &$21.49\pm0.07$ &$20.86\pm0.03$  &$19.46\pm0.04$ &$18.47\pm0.03$ &$\cdots$       &$\cdots$       &$\cdots$       &$\cdots$       &$\cdots$\\
2008-05-04 & 54590.4  &$21.52\pm0.08$ &$20.91\pm0.03$  &$\cdots$       &$18.48\pm0.03$ &$\cdots$       &$\cdots$       &$\cdots$       &$\cdots$       &$\cdots$\\
2008-07-05 & 54652.4  &$22.72\pm0.07$ &$22.27\pm0.03$  &$21.16\pm0.04$ &$20.03\pm0.04$ &$\cdots$       &$\cdots$       &$\cdots$       &$\cdots$       &$\cdots$\\
2008-07-18 & 54666.2  &$\cdots$       &$\cdots$        &$\cdots$       &$\cdots$       &$\cdots$       &$\cdots$       &$\cdots$       &$14.97\pm0.01$ &$14.04\pm0.01$\\
2008-11-22 & 54792.1  &$\cdots$       &$23.59\pm0.05$  &$22.50\pm0.05$ &$\cdots$       &$\cdots$       &$\cdots$       &$\cdots$       &$\cdots$       &$\cdots$\\
2008-11-23 & 54793.1  &$\cdots$       &$23.58\pm0.06$  &$22.56\pm0.05$ &$\cdots$       &$\cdots$       &$\cdots$       &$\cdots$       &$\cdots$       &$\cdots$\\
2008-11-24 & 54794.1  &$\cdots$       &$23.45\pm0.05$  &$22.45\pm0.05$ &$\cdots$       &$\cdots$       &$\cdots$       &$\cdots$       &$\cdots$       &$\cdots$\\
2008-11-25 & 54795.1  &$\cdots$       &$23.54\pm0.06$  &$22.60\pm0.05$ &$\cdots$       &$\cdots$       &$\cdots$       &$\cdots$       &$\cdots$       &$\cdots$\\
2009-03-25 & 54915.5  &$< 24.1$       &$< 25.6$        &$< 24.8$       &$23.10\pm0.07$ &$\cdots$       &$\cdots$       &$\cdots$       &$\cdots$       &$\cdots$\\
2009-10-20 & 55124.1  &$< 25.2$       &$< 25.9$        &$< 25.7$       &$< 25.1$       &$\cdots$       &$\cdots$       &$\cdots$       &$\cdots$       &$\cdots$\\
2009-10-22 & 55126.1  &$< 24.9$       &$< 25.9$        &$< 25.6$       &$< 25.1$       &$\cdots$       &$\cdots$       &$\cdots$       &$\cdots$       &$\cdots$\\
2009-12-17 & 55182.0  &$\cdots$       &$\cdots$        &$\cdots$       &$\cdots$       &$\cdots$       &$20.31\pm0.14$ &$\cdots$       &$\cdots$       &$\cdots$\\
2010-03-17 & 55272.5  &$\cdots$       &$\cdots$        &$\cdots$       &$\cdots$       &$\cdots$       &$< 21.4$       &$19.23\pm0.09$ &$\cdots$       &$\cdots$\\
2010-03-18 & 55237.5  &$< 24.6$       &$<25.3$         &$<25.4$        &$< 24.9 $      &$\cdots$       &$\cdots$       &$\cdots$       &$\cdots$       &$\cdots$\\
2010-05-17 & 55333.4  &$\cdots$       &$\cdots$        &$\cdots$       &$\cdots$       &$\cdots$       &$\cdots$       &$20.27\pm0.15$ &$\cdots$       &$\cdots$\\
2010-08-08 & 55417.3  &$\cdots$       &$\cdots$        &$\cdots$       &$\cdots$       &$\cdots$       &$\cdots$       &$\cdots$       &$15.91\pm0.02$ &$14.66\pm0.01$\\
2010-08-24 & 55433.2  &$\cdots$       &$\cdots$        &$\cdots$       &$\cdots$       &$< 25.9$       &$23.00\pm0.03$ &$\cdots$       &$\cdots$       &$\cdots$\\
2010-12-01 & 55531.9  &$\cdots$       &$\cdots$        &$\cdots$       &$\cdots$       &$\cdots$       &$\cdots$       &$\cdots$       &$16.16\pm0.02$ &$14.86\pm0.01$\\
2011-08-07 & 55780.6  &$\cdots$       &$\cdots$        &$\cdots$       &$\cdots$       &$< 26.5$       &$24.78\pm0.21$ &$\cdots$       &$\cdots$       &$\cdots$\\
\tableline
\end{tabular}
\tablenotetext{}{All the magnitude upper limits are $3\sigma$. The
  estimated start date of the transient is MJD $54485.5 \pm 4$
  \citep{Botticella2009}. $B$, $V$ and $R$ are Bessel filters, $Us$ is
  a high throughput $U$-band interference filter.}
\end{center}
\end{table*}

\begin{figure*}[htb]
\centerline{\includegraphics[width=12cm]{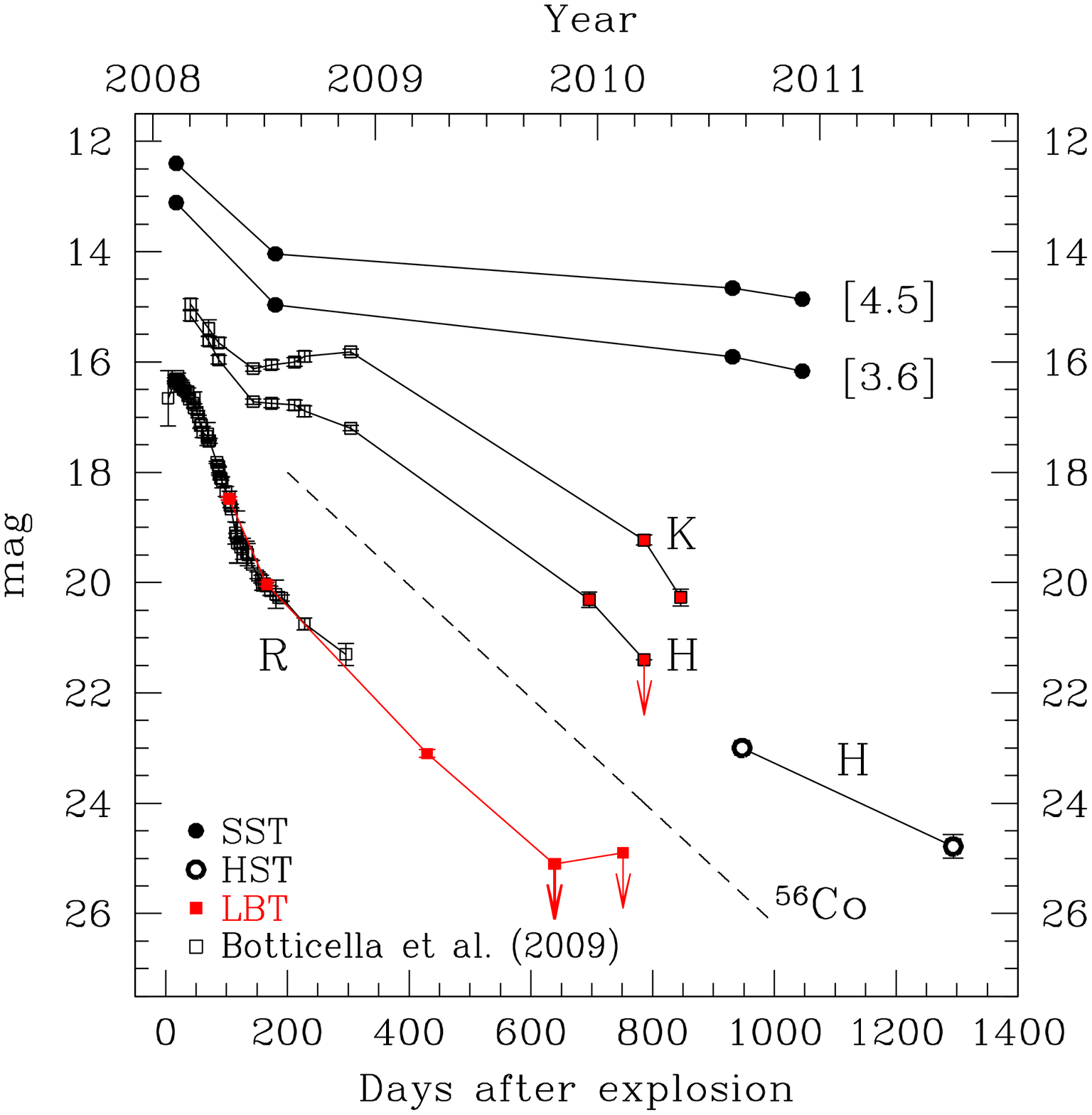}}
\caption{The $R$, $H$ and $K$-band light curves of SN~2008S from
  Botticella et al. (2009, open black points),the Large Binocular
  Telescope (filled red points) and HST (open circles). Filled
  circles show the $3.6$ and $4.5 \mu$m Spitzer light curves. Upper
  limits are indicated by arrows.  \cite{Botticella2009} noted that
  the early (<200 days) bolometric light curve had a slope consistent
  with that for $^{56}$Co decay (indicated by the dashed line), but
  this is clearly inconsistent with the late time light curve.
}
\label{fig:lc}
\end{figure*}

The only means of having an optical eruption from a massive star and
an invisible progenitor is for the star to be self-obscured by dust
that is largely destroyed by the transient.  This possibility was
confirmed when \cite{Prieto2008} found the progenitor star as a $\log
L/L_\odot \simeq 4.5$, $T\simeq 440$~K blackbody in archival Spitzer
data. This luminosity is comparable to that of an evolved $\sim 10\,
M_\odot$ star, and is well below that corresponding to the more
massive stars thought to be required for non-supernova eruptions. 
Subsequent analyses of the progenitor by \cite{Botticella2009} and
\cite{Wesson2010} were consistent with those by \cite{Prieto2008}.

More remarkably, an almost identical event then occurred in NGC~300
(\citealt{Monard2008}).  The progenitor was invisible in the optical
to even tighter limits (\citealt{Berger2008}; \citealt{Bond2009};
\citealt{Berger2009}), but we again found the progenitor as a
self-obscured star of similar luminosity and (dust photosphere)
temperature in Spitzer mid-IR data (\citealt{Prieto2008a};
\citealt{Thompson2009}).  A subsequent analysis of the progenitor by
\cite{Berger2009} agreed with our estimates, and an investigation of
the progenitor based on its neighboring stars by \cite{Gogarten2009}
was consistent with the progenitor being a massive star of order
$10$-$20\,M_\odot$, where the analysis favored the upper portions of
this range but, strictly speaking, the method only provides an upper
mass bound.

In \cite{Thompson2009} we surveyed the galaxy M33 for mid-IR sources
with similar properties to these progenitors and found that they were
astonishingly rare, with only a few such sources in the entire galaxy.
In the mid-IR, these sources have the properties of super-AGB stars,
with properties distinct from other classes of massive stars such as
LBVs and red supergiants. The rarity of these sources compared to all
massive stars, confirmed in our survey of additional galaxies
\citep{Khan2010}, means that the progenitors of the transients are a
very short lived ($\sim 10^4$~years) phase in the evolution of these
massive stars and that there is a causal connection between
obscuration and explosion.

\cite{Thompson2009} concluded that there are a number of possible
mechanisms to explain the nature of these transients and their
progenitors: (1) massive white-dwarf birth; (2) electron-capture
supernova; (3) intrinsically low-luminosity iron core-collapse
supernova; and (4) massive star outbursts. Debates about these
possible origins have been raging ever since then, based both on
theoretical and observational arguments.  They are basically divided
into the (some kind of) supernova camp (\citealt{Prieto2008};
\citealt{Botticella2009}; \citealt{Pumo2009}) and the (some kind of)
massive star outburst camp (\citealt{Berger2009}; \citealt{Smith2009};
\citealt{Bond2009}; \citealt{Kashi2010}; \citealt{Humphreys2011}).
The outburst camp generally
argues that the progenitor was not a $\sim 10\,M_\odot$ super-AGB star
but a more massive $15-20\,M_\odot$ star (supported by
\citealt{Gogarten2009}), despite their position at the red, high
luminosity end of the AGB sequence in mid-IR color-magnitude diagrams
(\citealt{Thompson2009}; \citealt{Khan2010}) and the low mass compared
to typical stars with LBV outbursts (see \citealt{Smith2010}). The
massive-star outburst interpretation is seriously called into question
by our Spitzer IRS spectrum of the NGC~300 event (\citealt{Prieto2009},
also \citealt{Kwok2011}). The mid-IR spectrum resembles that of
carbon-rich proto-planetary nebulae and lacks the silicate-dominated
dust features typical of massive star outbursts (e.g.,
\citealt{Humphreys2006}).  \cite{Wesson2010}, analyzing post-event
Spitzer observations of SN~2008S, also found that the silicate dust
characteristics of high mass stars were inconsistent with the
observations.  \cite{Prieto2009} also note that proto-planetary
nebulae (initial masses $ \ltorder 8\,M_\odot$) have most of the
optical spectral features that led \cite{Smith2009}, \cite{Bond2009}
and \cite{Berger2009} to argue for an outburst from a more massive
($\sim 20\,M_\odot$) star.  Since ``Type IIn'' optical spectroscopic
properties are seen in some proto-planetary nebulae, massive
supergiants, supernova impostors, and the genuine, but very diverse,
Type IIn supernovae, they appear only to be a diagnostic for the
presence of strong interactions between ejecta and a dense
circumstellar medium rather than a diagnostic for the source of the
ejecta.

Most recently \cite{Kochanek2011} re-analyzed the available data for
SN~2008S and NGC~300-OT and concluded that both transients were of 
explosive nature. If the transients were not explosive, the energy
would not be sufficient to destroy the obscuring dust and make the
transients visible in the optical. After the peak of the transient,
the dust rapidly reforms from the outside, and the luminosity, which
then comes from the shock propagating through the dense wind, is
mostly reradiated by that dust in the IR. The model predicts that the
peak obscuration is 2-3 years post-transient, when the shock reaches
the reformed dust shell and pushes its radius outwards. It should take
another 5-10 years (in the case of SN~2008S) before it is possible to
see through the dust and determine the fate of the progenitor star.

Here we report the data on SN~2008S used in this analysis. We have
been following the SN~2008S event with the LBT in both the optical
and near-IR, with HST in the near-IR, and with SST in the mid-IR.
Here we report that the source is again too faint to detect in the
optical, marginally detected in near-IR HST images (see
Fig.~\ref{fig:hst}), and still bright in the mid-IR. It is presently
much more luminous than the progenitor and slowly fading.
We describe our observations and results in \S2 and discuss their
implications in \S3.

\begin{figure*}
\centerline{\includegraphics[width=18cm]{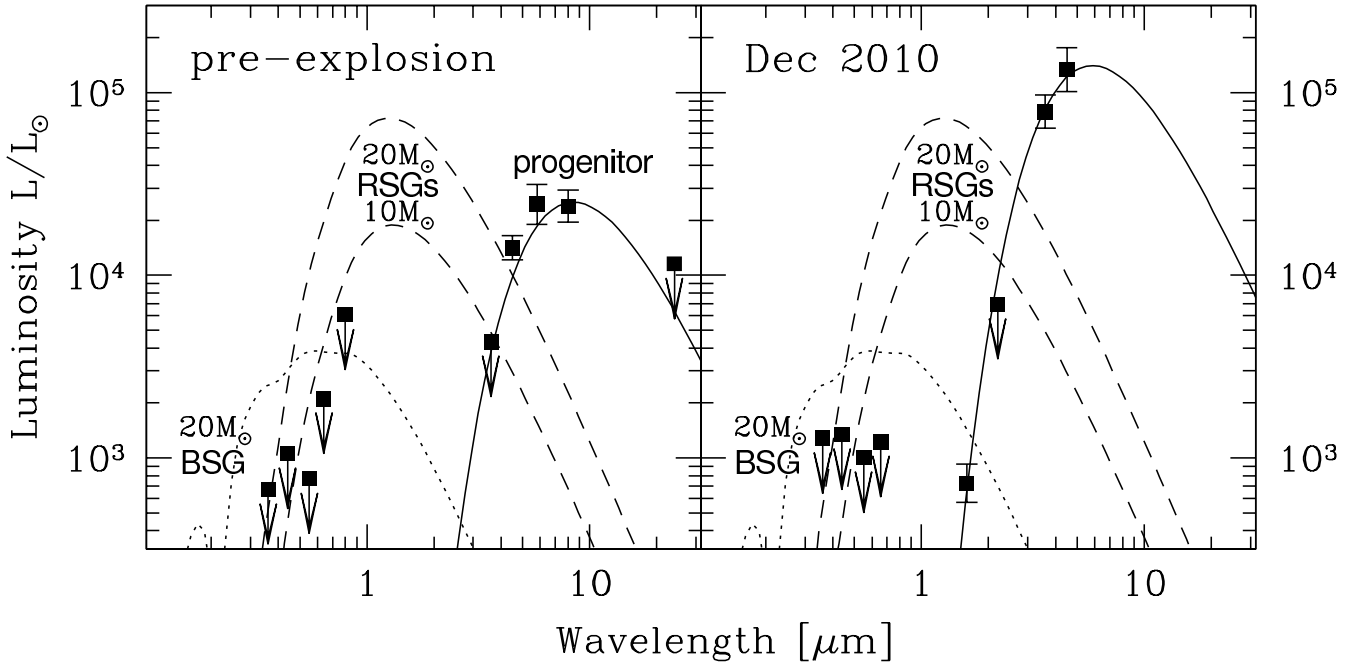}}
\caption{The pre-explosion, progenitor SED (left) and the current SED
  (right) of SN~2008S as of December 2010. The measured magnitudes are
  converted to fluxes, and these are converted to a luminosity as
  $L=4\pi D^2 \nu  F_\nu$ where $D=5.6$~Mpc. The SED models are just
  blackbodies plus $A_V=2.13$~mag of total extinction. The $10\,M_\odot$
  and $20\,M_\odot$ red supergiant models (RSG, dashed curves) are
  from Marigo et al. (2008) and have $T \simeq 3600$ and $3900$~K
  with $\log L/L_\odot = 4.68$ and $5.29$, respectively.  The blue
  supergiant model (dotted curve) is based on SN1987A and has $T
  \simeq 16000$~K and $\log L/L_\odot = 5.0$.  The best fit blackbody
  model (solid curve) for the progenitor has $T=440$~K and $\log
  L/L_\odot=4.54$ \citep{Prieto2008} while the December 2010 model is
  consistent with $T=640$~K and $\log L/L_\odot=5.30$. The last
  available LBT $K$-band observation (see Table~\ref{tab:magnitudes})
  is used as an upper limit on the right panel.}
\label{fig:sed}
\end{figure*}

\section{Observations and Results}

The optical observations were done with the Large Binocular Cameras
(LBC, \citealt{Giallongo2008}), using the LBC/Blue camera for $U$, $B$
and $V$ and the LBC/Red camera for $R$.  The pixel scale of the LBC
cameras is $0\farcs224$.  Since these observations are part of a
program whose overall goal is to use difference imaging to
characterize variable sources, the sub-images obtained for each epoch
were not dithered and SN~2008S was always located at approximately the
same point on Chip~2 of the cameras. Image exposure times were
$300$~sec, generally with two exposures for $U$, $B$ and $V$ and 6
exposures for $R$.  The near-IR observations were made with LUCIFER
(\citealt{Seifert2003}; \citealt{Mandel2008}; \citealt{Ageorges2010})
in the $H$ and $K$ bands using the F3.75 camera with a pixel scale of
$0\farcs12$.  At each dither position we obtained 3 exposures of 33
(10) sec for $H$ ($K$) band.  We obtained 10 on-source and 6
off-source dither positions in a 2-5-2-5-2 off-on-off-on-off pattern,
where the off-source position was shifted 8~arcmin away from the
galaxy. Another set of near-IR observations was obtained with the
Hubble Space Telescope (HST) WFC3/IR camera in $F110W$ ($J$) and
$F160W$ ($H$) filters (proposal ID 12331). The exposure times were
1x800s in $J$ and 3x700s in $H$ and were taken in August 2010 and 2011.
In addition to the near-IR HST observations, we obtained mid-IR
data with warm Spitzer on August and December 2010 (program ID 70040),
with exposure times of 8$\times$30s at both $[3.6]$ and $[4.5]$.

The optical and near-IR LBT data were reduced using standard methods
in IRAF. The photometry was obtained using DAOPHOT and ALLSTAR
(\citealt{Stetson1987}; \citealt{Stetson1992}).  The optical data was
calibrated using $4-24$ local standards from \cite{Welch2007} for the
$V$ and $R$ bands and from \cite{Botticella2009} for the $U$ and $B$
bands. The near-IR LBT data were calibrated using $3-6$ 2MASS stars in
the field. In both cases we only applied a zero-point offset to
convert the instrumental magnitudes into the standard system.
The photometry on the HST near-IR data was made with the DOLPHOT
package (\citealt{Dolphin2000}) utilizing the WFC3 module which
provides specific PSFs and pixel area maps and automatically applies
all the necessary corrections. For the mid-IR SST data, we used
aperture photometry with an aperture radius 4 pixels (2$\farcs$4) and
a background annulus 4-12 pixels. We applied the aperture corrections
of 1.213 for $[3.6]$ and 1.234 for $[4.5] \mu$m provided in the SST
IRAC Instrument Handbook.\footnote{http://irsa.ipac.caltech.edu/data/SPITZER/docs/irac/iracinstrumenthandbook/28/}
The multi-wavelength results are presented in
Table~\ref{tab:magnitudes}, where the magnitude errors include the
uncertainties both in the measurements and in the zero points. In the
cases where we do not detect SN~2008S, we place a $3\sigma$ upper
limit on the magnitude using the standard deviation of the sky in a
region around the source.

Figure~\ref{fig:lc} shows the $H$, $K$ and $R$-band light curves from
Botticella et al. (2009) and our LBT observations, as well as the
$H$-band HST and $[3.6]$/$[4.5] \mu$m SST data points. The SN is not
detected in the HST $J$-band and we do not plot the limits for clarity.
For the same reason we do not plot the most recent $R$-band LBT limits.
Figure~\ref{fig:sed} shows the progenitor and current (as of December
2010) spectral energy distributions (SED).
The left panel of Fig.~\ref{fig:sed} shows the constraints on the
progenitor's SED as compared to typical massive stars. To make the
comparison we used a Galactic plus intrinsic extinction of
$A_V=2.13$~mag \citep{Botticella2009} and the distance of $D=5.6$~Mpc
adopted by \cite{Prieto2008}.  The data points are converted to a
luminosity as $L=4\pi D^2 \nu F_\nu$.  For comparison we show the
extincted SEDs of $10\,M_\odot$ and $20\,M_\odot$ red supergiants
(RSG) using luminosities and effective temperatures from
\cite{Marigo2008}, a $20\,M_\odot$ blue supergiant (BSG) modeled on
SN1987A, and the blackbody that best fit the SN~2008S progenitor data
(solid line).

In the optical ($UBVR$), the source is again too faint to correspond
to a massive ($>10 \,M_\odot$) evolved star, with limits on its
brightness similar to those for the progenitor (see right panel in
Fig.~\ref{fig:sed}). The extinction would have to be increased from
the $A_V\simeq 2.1$~mag estimated to be present post-explosion
(\citealt{Botticella2009}) to $A_V\sim 3.6-5.8$~mag in order to
obscure the models shown in Fig.~\ref{fig:sed}. The transient is
now marginally detectable in the HST near-IR (see the middle panel of
Fig.~\ref{fig:hst}), and it is fading with a slope of
$1.9 \pm 0.25$~mag/year between the 2 HST epochs which is less steep
than the mean slope of $2.9 \pm 0.2$~mag/year between the late phases
of the \cite{Botticella2009} $H$-band light curve and our first LBT
observation. The decline at $K$-band between the last two LBT epochs
is very rapid, approximately $6 \pm 1$~mag/year and is significantly
steeper than the mean slope of $2.3 \pm 0.1$~mag/year between the
\cite{Botticella2009} and LBT light curves.  The SED is rising to the
red with $H-K > 2.2$~mag 800 days after explosion.  If we extrapolate
the $H$-band flux from December 2009 to March 2010 (which is the last
$K$-band data point) by connecting the last LBT and first HST
detections, we estimate $H \simeq 21.9$~mag and thus
$H-K \simeq 2.7$~mag, which is significantly redder than the
$H-K \simeq 1.4$~mag color in the late phases of \cite{Botticella2009}.

We can roughly estimate a temperature and luminosity for the March
2010 epoch.  Fitting a blackbody to the measured $K$-band flux and
the extrapolated $H$-band estimate ($21.9$~mag), we get a temperature
of 700~K and a luminosity of $280000\,L_\odot$. When we estimate
a temperature and luminosity for the last epoch of SST observations
(December 2010) we get $T=640$~K and $L=200000\,L_\odot$. In the DUSTY
(\citealt{Ivezic1997}) models of these data by \cite{Kochanek2011},
the estimated luminosity at this epoch is $\log L/L_\odot = 5.6 \pm 0.6$.
Even with the further fading between March and December 2010, the
source luminosity is still much higher than the estimated luminosity
$L \simeq 40000\,L_\odot$ of the progenitor star (\citealt{Prieto2008};
\citealt{Botticella2009}; \citealt{Wesson2010}).
The source has faded by another 1.25~mag in the $H$-band between
December 2010 and August 2011 and by now is probably undetectable in
the near-IR. Future warm Spitzer observations will allow us to follow
the mid-IR evolution of the light curve, but without cold Spitzer
data the SED will be hard to constrain.

\begin{figure*}
\includegraphics[width=18cm]{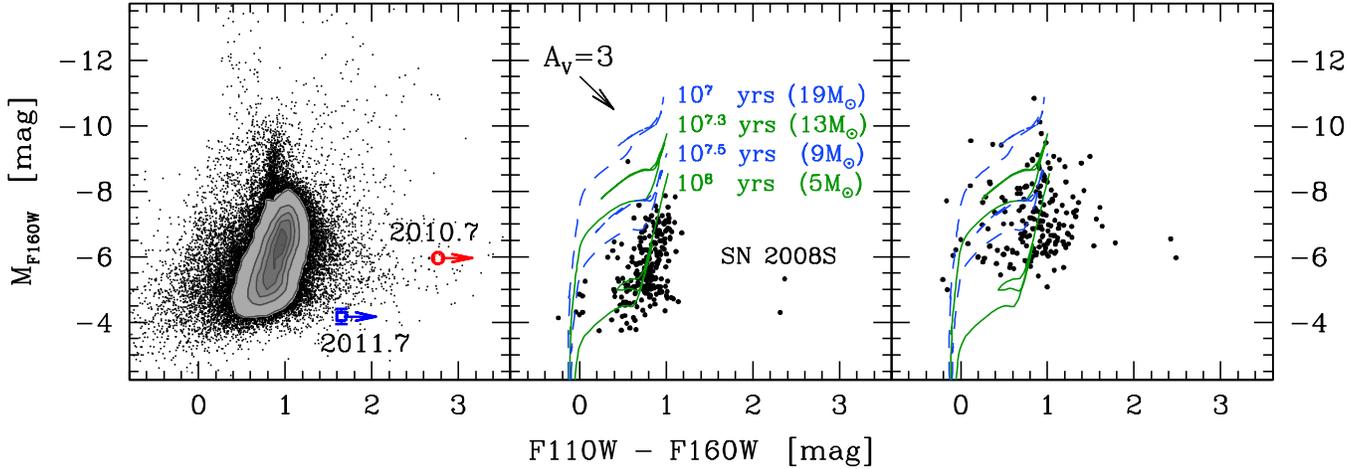}
\caption{Near-IR CMDs from the combined HST epochs. The left panel
shows a large region (2.3$\times$2.0 arcmin) while the middle and
right panels show a 4\farcs1 (100 pc) radius around SN~2008S and
SN~2002hh, respectively.
The CMDs are corrected for Galactic extinction and assume a distance
of $5.6$~Mpc.  The location of SN~2008S is marked with a circle for
the 2010 epoch and with a square for the 2011 epoch. The curves on the
middle and right panels show Padova (\citealt{Marigo2008}) isochrones
for $10^7$, $10^{7.3}$, $10^{7.5}$ and $10^{8}$~years, alternating
solid and dashed lines, which have end points corresponding to ZAMS
masses of $19$, $13$, $9$ and $5M_\odot$ respectively. The top of the
two reddest points in the middle panel is the epoch-averaged
measurement for SN~2008S.
}
\label{fig:cmd}
\end{figure*}

Fig.~\ref{fig:cmd} shows near-IR $H$/$J-H$ color magnitude diagrams
(CMD) for both a large (2.3$\times$2.0 arcmin) and a small 4\farcs1
(100 pc) radius region around the source (this smaller region is shown
in Fig.~\ref{fig:hst}).  These assume a distance of $5.6$~Mpc,
following \cite{Prieto2008}, and are corrected for $E(B-V)=0.342$~mag
of foreground Galactic extinction (\citealt{Schlegel1998}).
To construct the CMDs we combined the 2010 and 2011 HST epochs for
each band and then used DOLPHOT to perform photometry on the combined
images. The resulting photometric catalogs contained 82,000 sources.
Following \cite{Dalcanton2011}, we then required the signal-to-noise
ratio in both filters to be greater than 4 for a detection, and a
sharpness parameter $sharpness^2 < 0.1$ to exclude non-stellar sources
from the catalog, which left 67,000 sources. The resulting CMD is
dominated by old ($\gtrsim 1$ Gyr) stellar populations (see Fig.~10
from \cite{Dalcanton2011}). The red giant branch (RGB) is visible at
$F110W - F160W \sim 1.0$, but the red clump (RC), usually found $3-4$
magnitudes below the tip of the RGB (at $M_H \simeq -6$), is too faint
to be resolved in our data, mainly because of the high crowding. The
location of SN~2008S is marked with a circle for the 2010 epoch and
with a square for the 2011 epoch. It is very faint and red compared
to other stars in the field.

When we examine the CMD of the stars within a $4\farcs1$ (100~pc)
radius of SN~2008S in the middle panel of Fig.~\ref{fig:cmd}, we see
very few luminous stars.  The superposed Padova (\citealt{Marigo2008})
isochrones for $10^7$, $10^{7.3}$, $10^{7.5}$ and $10^{8}$~years have
ZAMS masses corresponding to their end points of $19$, $13$, $9$ and
$5M_\odot$, respectively.  Unless the local extinction is high, as
indicated by the arrow corresponding to the effect of $A_V=3$ of
additional extinction, the region lacks the population of young stars
that should be associated with a high mass ($M \gtorder 10 M_\odot$)
progenitor for SN~2008S, but does show evidence for a population with
$M \simeq 10 M_\odot$. For contrast, the right panel of
Fig.~\ref{fig:cmd} shows the CMD for  a $4\farcs1$ (100~pc) radius
region around the Type  IIP SN~2002hh (\citealt{Pozzo2006}). Here we
see a population of massive stars extending up to the $10^7$ yrs
isochrone, corresponding to a maximum mass of $19 M_\odot$, which is
well-matched to the upper mass limit of $M < 18 M_\odot$ for the
progenitor of SN~2002hh by \cite{Smartt2009}.

\section{Discussion}

\cite{Thompson2009} proposed that SN~2008S and the NGC~300 transient
were the archetypes of a new class of transients potentially including
the M85 OT-1 transient (\citealt{Kulkarni2007};
\citealt{Pastorello2007}), SN~1999bw (\citealt{Li2002} and references
therein), and now PTF10fqs (\citealt{Kasliwal2010}). The initial
defining characteristics were (1) a dust-enshrouded progenitor without
optical counterpart and mid-IR magnitudes that places them at the tip
of the AGB sequence in a mid-IR CMD, and (2) a low-luminosity
transient ($-13 \gtrsim M_{V} \gtrsim -15$) with narrow lines in
emission in the spectra ($v \lesssim 3000$~km/s), and signs of a
circumstellar dust excess at near-IR and mid-IR wavelengths.
Examinations of the dust properties (\citealt{Prieto2009};
\citealt{Wesson2010}) suggest (3) that the dust is carbonaceous rather
than the silicate dust seen in massive stars.

Here we add (4) that the progenitor either does not survive or must
return to its dust enshrouded state. As the right panel of
Fig.~\ref{fig:sed} shows, the LBT data already rule out the presence
of a massive, evolved star unless it has reconstituted an optically
thick, dusty envelope. The present optical limits are somewhat
stronger than those for the progenitor, and the near-IR detections
already rule out RSGs more massive than $10\,M_\odot$ unless more
heavily obscured. The total luminosity is still much higher than that
of the progenitor and emerges mainly in the mid-IR, but it is also
slowly fading.  \cite{Prieto2010} and \cite{Ohsawa2010} find that
the NGC~300 transient has also vanished in the optical but remains
bright in the mid-IR.

In the \cite{Kochanek2011} scenario, the shock powering the present
luminosity should eventually destroy the dust, and we will be able
to observe the direct emission from the shock and any surviving star.
This will likely require, however,
monitoring these sources for almost a decade.  These observations,
and the similar data for the NGC~300 transient, demonstrate, however,
that it is impossible to understand these transients based only on
optical observations -- understanding their evolution, luminosities, 
and energetics requires mid-IR observations.

The observational properties of the progenitors are broadly 
consistent with being massive ($M \simeq 10 M_\odot$) AGB stars.
The one significant counterargument has been the observation by
\cite{Gogarten2009} that the stellar populations near the NGC~300 
transient are consistent with a progenitor mass as high as 
$\sim 20 M_\odot$.  This approach technically only yields an upper
bound on the mass -- allowing, but not requiring, the NGC~300
progenitor to be significantly more massive than any AGB star. For
the neighborhood of SN~2008S, we see no evidence for a population
of stars significantly more massive than $\sim 10 M_\odot$ unless
there is an enormous amount of (uniform) local extinction.  This is
based, however, on a near-IR CMD in which it is difficult to constrain
extinctions and temperatures. Adding optical observations would make
it clear whether there is any possibility of younger more massive
stars in the region.

\acknowledgements 

We thank G. Cresci, J. Hill, R. Humphreys, and A. Quirrenbach for
suggestions and comments. 
JLP acknowledges support from NASA through
Hubble Fellowship grant HF-51261.01-A awarded by STScI, which is
operated by AURA, Inc. for NASA, under contract NAS~5-2655. CSK, JLP,
KZS, DS and TAT are supported in part by NSF grant AST-0908816. JFB is
supported by NSF CAREER Grant PHY-0547102 and NSF Grant PHY-1101216.
Based in part on observations made with the
Large Binocular Telescope. The LBT is an international collaboration
among institutions in the United States, Italy and Germany. The LBT
Corporation partners are: the University of Arizona on behalf of the
Arizona university system; the Istituto Nazionale di Astrofisica,
Italy; the LBT Beteiligungsgesellschaft, Germany, representing the Max
Planck Society, the Astrophysical Institute Potsdam, and Heidelberg
University; the Ohio State University; and the Research Corporation,
on behalf of the University of Notre Dame, University of Minnesota and
University of Virginia. 
This work is based in part on observations made with the Spitzer Space
Telescope, which is operated by the Jet Propulsion Laboratory, California 
Institute of Technology under a contract with NASA. Support for this work was 
provided by NASA through award 1414623 issued by JPL/Caltech.
Support for HST program GO-12331 
was provided by NASA through a grant from the Space Telescope
Science Institute, which is operated by the Association
of Universities for Research in Astronomy, Inc., under
NASA contract NAS5-26555.

{\it Facilities:} \facility{LBT, HST, SST}

\end{document}